# Development of the RFQ Cooler SHIRaC: beam transport and nuclearization

**Ramzi Boussaid**[a*], **G.Ban** [a], **G.Quéméner** [a], **Y.Merrer** [a], **J.Lorry** [a].

[a] *LPC-IN2P3, ENSICAEN, 6 Boul. Maréchal Juin, 14050 Caen, France.*

*E-mail*: boussaidramzii@gmail.com

**ABSTRACT;**

The development of the new RFQ Cooler, called SHIRaC, was carried out. As a part of SPIRAL 2 facility, SHIRaC aims to handle and cool typical SPIRAL 2 beams with large emittances (up to 80 π.mm.mrad) and high currents (up to 1 µA). Its purposes are to enhance as much as possible the beam quality (transverse geometric emittance of less than 3 π.mm.mrad and longitudinal energy spread close to 1 eV) and to transmit more than 60 % of ions.

Numerical simulations and experimental studies have shown that the required beam quality can be reached only in term of the emittance. The energy spread is very far from expected values. It is sensitive to the space charge and the buffer gas diffusion and more importantly to the RF field derivative effect. The latter arises at the RFQ exit and increases with the RF parameters (the frequency and the amplitude of the RF voltage).

Studies allowing to enhance the cooled beam quality, mainly the energy spread reduction, are presented and discussed along this paper. They consist in implementing a miniature RFQ at the RFQ exit. Using this method, it becomes possible to improve the cooled beam quality and to reach 1 eV of longitudinal energy spread and around 1.75 π.mm.mrad of transverse geometric emittance for beam currents going up to 1 µA.

The transport of the cooled beam from SHIRaC towards a HRS has been done with an electrostatic quadrupole triplet. Simulations and first experimental tests showed that more than 95 % of cooled beams can reach the HRS.

Finally, developments related to the nuclearization protection methods aiming to avoid the escape of any nuclear matter from the SHIRaC beamline are studied.

# Contents

# 1. Introduction

In the context of the new nuclear facility generation for the production of rare and exotic ion beams with intensities up to 1µA and emittances up to 80 $\pi$.mm.mrad [1], SPIRAL-2 project is currently installed at GANIL laboratory in France [2, 3, 4].

SPIRAL-2 is devoted to providing beams to DESIR (**D**esintegration, **E**xcitation and **S**torage of **I**on **R**adioactive) experiment [5, 6]. Such beams will be produced by the ISOL method and subsequently they will suffer from isobaric contaminations [7]. As DESIR's typical experiment requires the highest purified beam, a HRS (**H**igh **R**esolution **S**eparator) will be installed in front of DESIR [8, 9]. The nominal HRS working requires beams with low beam quality: transverse geometric emittance $\leq 3$ $\pi$.mm.mrad and longitudinal energy spread $\Delta E \sim 1$ eV [8]. In order to make the required purification with the highest efficiency, the beams should be cooled before reaching the HRS entrance. The **R**adio-**F**requency **Q**uadrupole **C**ooler (RFQC) is widely known as a universal cooling technique of ion beams [10].

The RFQCs are already used in several projects [11, 12, 13, 14 ]. The existing ones can only handle beams with low intensities (~100 nA) and small emittances (~10 π.mm.mrad). As the SPIRAL-2 typical beams are of higher currents and larger emittances, a new RFQC called SHIRaC (**S**piral-2 **H**igh **I**ntensity **Ra**diofrequency **C**ooler) is developed and tested at LPC-Caen laboratory in France [15, 16, 17]. In addition to its new optics system which captures the largest beam emittance, its vacuum system and RF system were also developed [15]. The vacuum system, based on a differential pumping, allows to reduce the buffer gas pressures to less than 0.01 Pa beyond the RFQ chamber. Whereas, with the developed RF system, highest RF parameters of up to 9 MHz and 9 kV of frequency and amplitude, respectively, can be reached without any breakdown limitations. These RF parameters provide the needed RF confinement field to overcome the space charge effects, due to the high beam currents. This Cooler is devoted to provide the beam quality required by the HRS as well as to transmit more than 60 % of incoming ions toward the HRS.

Simulation [16] and experimental studies [15] of SHIRaC prototype has been investigated and showed pertinent results. The prototype allows to handle and cool beams of currents going up to 1 µA which has never been achieved before. Using RF voltage parameters of 4.5 MHz in frequency and 4 kV in amplitude (Mathieu parameter q=0.4) and buffer gas pressure of 2.5 Pa, cooled beam quality with less than 2.5 π.mm.mrad of emittance and less than 6.5 eV of ΔE for above 70 % of incoming ions are provided by this prototype. The cooled beam quality is in agreement with the HRS requirements only in the term of emittance. The second beam quality term is much greater than the expected values and its reduction is mandatory for an optimum running of the HRS.

These studies also showed that the degrading effects (the space charge, the RF heating and the buffer gas diffusion) have a considerable impact on the cooling process and thereafter on the beam quality [15, 16]. Another pertinent degrading effect may occur at the RFQ exit and acts to grow the ΔE. A description and analysis of the origin of this effect will be investigated in the first section of this paper.

Study and development of an engineering technique to reduce this pertinent effect and then to avoid the degradation of the cooled beam quality beyond the RFQ exit are reported. Additional special considerations to reach cooled beam quality with ΔE ~ 1 eV and 1 π.mm.mrad of emittance with higher ion transmission are investigated.

In order to transport the cooled beam from the extraction section exit toward the HRS entrance, an electrostatic quadrupole triplet was developed. More details about the simulation studies and experimental tests of the triplet are also presented.

Our purpose to study protection methods against the radioactivity effect, due to the high intensity of SPIRAL 2 typical ion beams, is also illustrated.

## 2. Physics motivation

As a recall, SHIRaC is composed of three separate sections: the injection section, the RFQ section and the extraction section [16]. These sections are connected by exchangeable apertures where the diffusion of the buffer gas feeding the RFQ chamber is governed by a differential pumping system.

Following the cooling through the RFQ section, the cooled beam is extracted and then is guided with a small acceleration (only few eV) to the extraction plate exit. Once the beam passes through this plate, it is strongly accelerated [16]. The most critical point in the acceleration of the beam is the area between the RFQ exit and the extraction plate exit, where the pressure is extremely high, close to the RFQ pressure [15]. The good beam quality achieved by the cooling could be partially lost due to undesirable collisions of the extracted ions with the buffer gas atoms. The problem worsens in the absence of a confinement field. In addition, the space charge will result in significant degradations as long as the ions' energy is very low.

Comparative simulation results of the beam quality [16], at the RFQ exit and at the extraction section exit, versus the beam currents showing these degradations, are presented in figure 1. Such degradations can be estimated by relying on the relative differences between the beam quality at the RFQ exit and the beam quality at the extraction section exit. Degradations stemming from the buffer gas diffusion correspond to the gap between the curves at low beam current, less than 100 nA. These degradations revolve around the emittance and more importantly around the energy spread and are worth 0.2 $\pi$.mm.mrad and 2.5 eV, respectively. The gap widens progressively with the beam currents while the space charge effect increases with them. The arising of this phenomenon can be observed in the quick degradation of the emittance with the beam currents (figure 1-left). The energy spread is in the same way affected as it increases (figure 1-right). Its degrading behavior is also observed experimentally as seen on figure 2. We also note a good agreement between the simulation and experimental results.

The occurred degradation at low beam current (less than 100 nA), where the space charge effect can be neglected, is expected to be due to the buffer gas diffusion and is of 10 % for the emittance and more than 300 % for $\Delta E$. The energy spread degradation is too large to be stemming only from the buffer gas diffusion because, on one hand, it is much greater than those provided with the existing cooler, notwithstanding that the buffer gas pressure is at the same level and that the vacuum system is not as powerful as the SHIRaC's [13] and, on the other hand, the gas diffusion had a small effect on the beam energy that did not exceed 10 %.

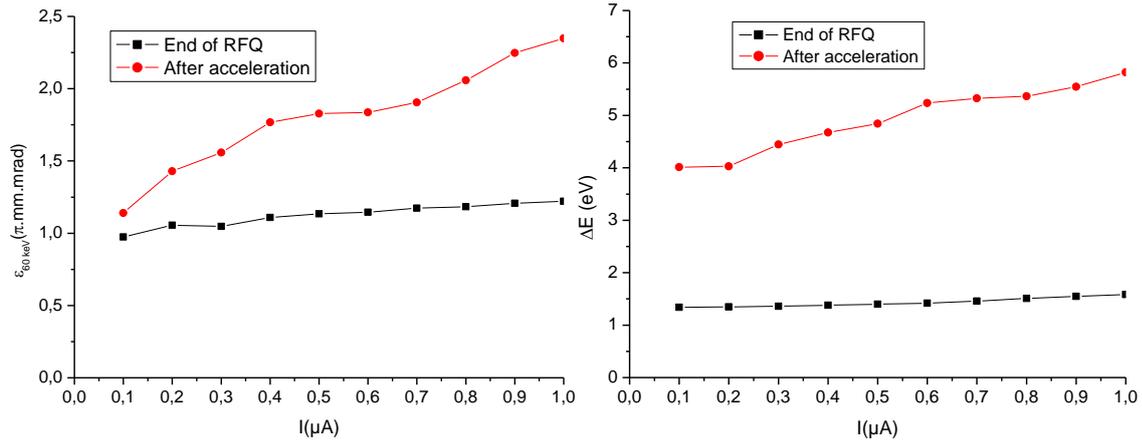

**Figure 1**: Space charge effect on the ion beam quality: simulation of the beam current effect on the beam emittance (right) and the longitudinal energy spread ΔE at the RFQ section exit and at the exit of the RFQ Cooler line (left) [16].

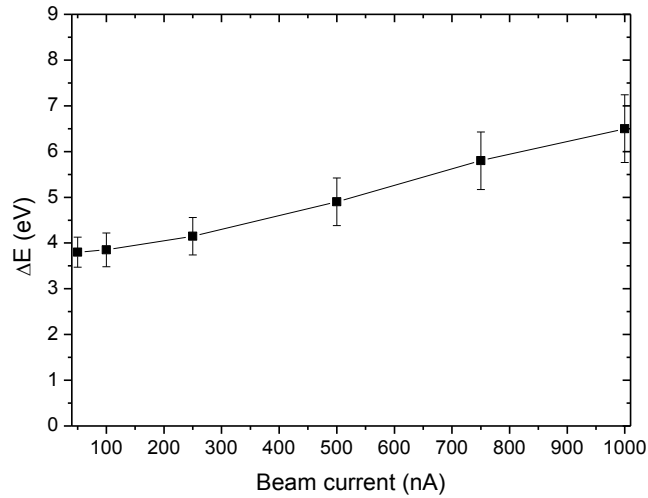

**Figure 2**: Space charge effect on the longitudinal spread energy: experimental results of the beam current effect on the longitudinal energy spread at the exit of the RFQ Cooler line [15].

The main difference between previous RFQCs and SHIRaC lies in the RF voltage parameters used, which are less than 1 kV and 1 MHz of RF voltage amplitude and frequency, respectively, in the former [18], compared to more than 2.5 kV and 4.5 MHz in the latter. This leads us to study the RF parameters effect on the energy spread. In figure 3, we present the variation of ΔE as a function of the RF voltage amplitude $V_{RF}$ for beam current of 1 µA. Contrary to expected behavior of cooled beam quality with the $V_{RF}$, an increase of the energy spread occurs. This increase is not explained by the effect of the RF voltage amplitude on the cooling (i.e., on the ΔE). Rather, the RF field derivative effect at the RFQ exit is used to explain it. The derivative of the transverse RF field at the RFQ exit gives rise to a longitudinal RF field which acts on the longitudinal velocities distribution of the cooled ions via its random characteristic. Thereafter, it results in a degradation of this distribution and, consequently, in growth of the energy spread.

The derivative of transverse RF field at the RFQ exit can possesses a transverse component. This component has the same confinement effect as the RF field into the RFQ, but with smaller power. This explains that the small degradation of the emittance with low beam current, which occurred after the RFQ exit (figure 1), is due to the buffer gas diffusion.

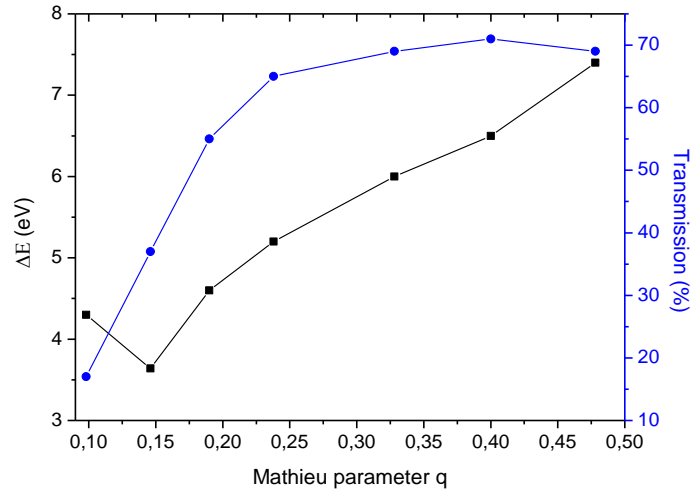

**Figure 3**: Confinement effect of the RF voltage amplitude on the longitudinal energy spread: variation of the longitudinal energy spread ΔE with the Mathieu parameter q (i.e RF voltage amplitude) [15].

## 3. Development of the extraction section

As cited above, the space charge and the buffer gas diffusion at the RFQ exit have a contribution in the energy spread growth. However, the RF field derivative effect has the greatest contribution in this degradation. In order to avoid this degradation, an optical device providing both the transport of cooled ions from one vacuum stage to another, with only minor disturbances, and a confinement field, should be implemented at the RFQ exit. One of the options is to set up a miniature RFQ (µRFQ).

### a. Miniature RFQ Cooler (µRFQ)

The µRFQ device technique was used in several projects [19, 20, 21] as a method allowing to just reduce the buffer gas diffusion. In the case of SHIRaC, this device should furthermore enable to:
- Overcome the space charge effects which occur at the RFQ exit.
- Reduce the RF field derivative effect: It acts as a cut-off of the RF voltage derivative propagation.
- Guide the cooled ion beam to a region where their energy is of a few tens eV, so then they can resist to any degrading effects.

Simulation studies of this µRFQ, related to the optimization of its design as well as its position relative to the RFQ end, have been done. These simulations intend to study, likewise, the cooled beam quality dependencies on the RF voltage amplitude $V_{RF}$, the buffer gas pressure $P_{RFQ}$ and the space charge. They are realized by SIMION 3D V8.1 program and the hard disc model as presented in reference [22, 23]. One thousand $_{133}Cs^+$ ion beams with 60 keV of energy and 80 π.mm.mrad of emittance were used. Preliminary simulations showed that the optimized µRFQ dimensions are 40 mm of length and 2 mm of inner radius. In order to keep the lowest buffer gas diffusion whilst avoiding any breakdown between the electrodes of the RFQ and those of the µRFQ, the µRFQ's electrodes are half cylinder. With this method, the pressure distribution beyond the RFQ chamber is reduced of about 10 %.

In order to avoid the RF field derivative effect and any acceleration or deceleration phenomenon of ions in the interface between the RFQ and the µRFQ, the latter must run with the same frequency as the RFQ's. For ions stability reasons along the µRFQ, low RF voltage amplitudes supplying their electrodes should be used. Thus, the µRFQ must operate with RF amplitude of 441 V for a frequency of 4.5 MHz, hence a Mathieu parameter q = 0.4.

In order to adequately extract the cooled beam from the µRFQ while avoiding any degrading effect on the obtained cooled beam quality, double extraction electrodes of 3 mm of inner radius are installed at the µRFQ exit (figure 4). The first electrode is of 1 mm from the µRFQ end and of 2 mm from the second electrode.

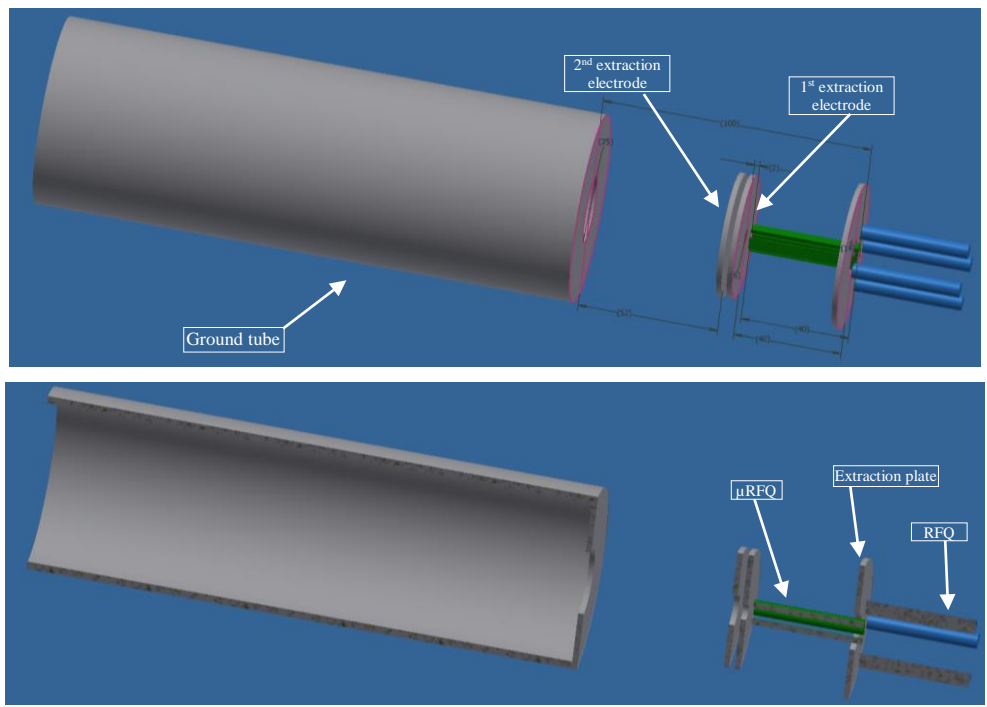

**Figure 4**: Layout of the developed extraction section and relative positioning of the µRFQ from the main RFQ and the.

## b. Longitudinal energy spread improvement

Using the new design of the extraction section, simulation results of the space charge and RF field derivative effects on the cooling process are illustrated in this section. The energy spread were investigated as a function of beam currents and RF voltage amplitudes.

The below figure illustrates simulated results of the ∆E as a function of the RF voltage amplitude (i.e., the Mathieu parameter q), applied to the main RFQ electrodes, while keeping constant the RF voltage amplitude (satisfying q= 0.4) applied to the µRFQ. The RF field derivative effect is clearly deleted because ∆E doesn't increase with the RF voltage amplitude, as seen in figure 5, but rather decreases until reaching a plateau for q between 0.3 and 0.5. The plateau mirrors the obtained optimum cooling. This behavior of ΔE reveals very clearly, on one hand, the expected RF voltage amplitude effect on the confinement power and thereafter on the beam cooling throughout RFQ section, and the cut-off of the field derivative effect by the µRFQ on the other hand. The increase in the ΔE for q > 0.5 is explained by the occurrence of the RF heating effect on the cooling.

Besides the removing of the derivative effect, we clearly note that both the space charge and the buffer gas diffusion effects are reduced in view of lower energy spread values.

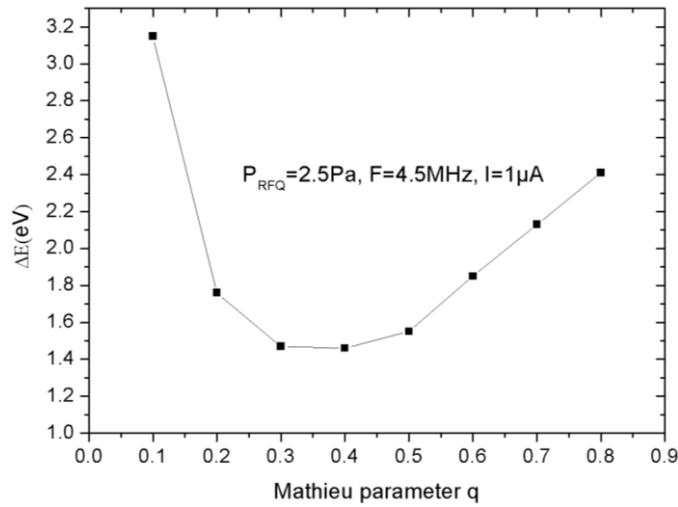

**Figure 5**: Effect of the RF voltage amplitude on the longitudinal energy spread ΔE for 1µA ion beam: variation of ΔE as a function of the Mathieu parameter q.

The µRFQ contribution to reduce the space charge effect on the energy spread should be noticeable in view of the presence of a confinement field to overcome this effect as long as the cooled ions are slow at the RFQ exit. This is very clear, on one hand, in figure 6-Left, as the ΔE increases little by little from 1.1 eV to 1.45 eV with beam currents going up to 1 µA. It is visibly manifested in the reduction of the variation gap of the ΔE, which was more than 2.5 eV without the µRFQ [16] versus only 0.3 eV with the present developments, for currents going from 50 nA to 1 µA. On the other hand, it is also clearly involved as the energy spread values are very smaller than those without the µRFQ, such as the ΔE that is reduced from 6.45 eV to 1.48 eV for 1 µA.

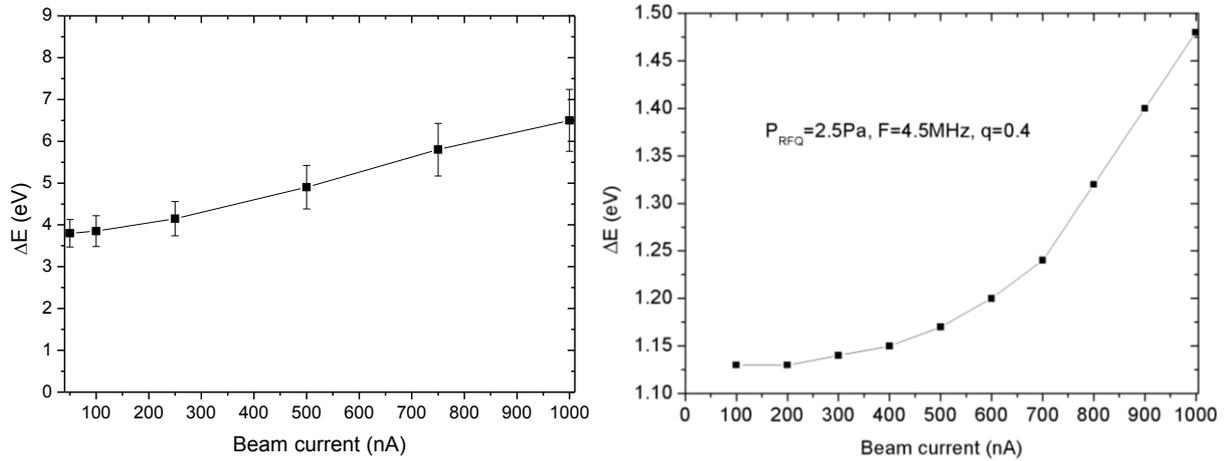

**Figure 6**: Illustration of the space charge effect on the longitudinal energy spread with and without the µRFQ and for Mathieu parameter q=0.4 and $P_{RFQ}$=2.5 Pa: Variation of ΔE as a function of the beam current without the µRFQ (left), Variation of ΔE as a function of the beam current in the presence of the µRFQ (right).

### c. Cooled beam quality improvement

As showed above, the µRFQ contribution is very efficient to reduce as low as possible the degrading effects of the cooled beam quality. A supplementary cooling of ions can be occurred in presence of the µRFQ. The cooling can be improved in increasing the buffer gas pressure while maintaining low buffer gas diffusion. Thus, any increase of RFQ pressure results in enhancing the cooled beam quality. The below figure illustrates this phenomenon as the minimum of ΔE, around 1.01 eV, is obtained with 4 Pa of pressure, instead of 2.5 Pa without the µRFQ. As is expected, ΔE is improved with the buffer gas pressure until reaching its optimum value at 4 Pa where the cooling is optimum. The increase in the ΔE at the high pressures, beyond 4.5 Pa, appeared to come from the occurrence of the buffer gas diffusion effect at the µRFQ exit.

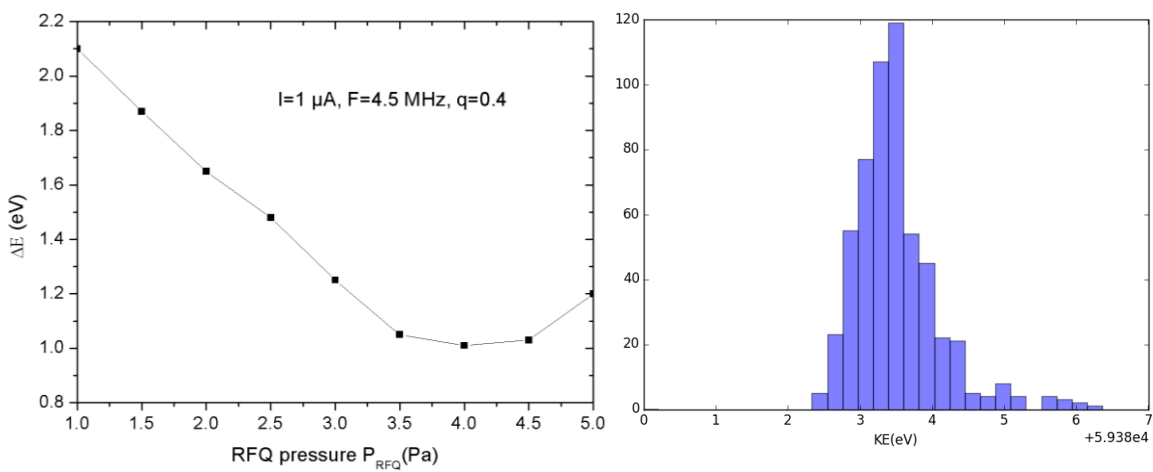

**Figure 7**: Effect of the RFQ pressure $P_{RFQ}$ on the longitudinal spread energy ΔE for 1 µA of beam current and q=0.4: variation of ΔE as a function of the $P_{RFQ}$ (left), beam energy histogram taken at 4Pa of $P_{RFQ}$(right).

The cooling improvement with the buffer gas pressure should automatically result in an emittance reduction which is very clear where the emittance decreases from 2.5 π.mm.mrad without the µRFQ [15] to around 1.75 π.mm.mrad respectively, with 2.5 and 4 Pa. It is also important to note that the simulations showed the transmissions remain unchangeable in implementing the µRFQ and exceed 70 % for beam current going up to 1 µA.

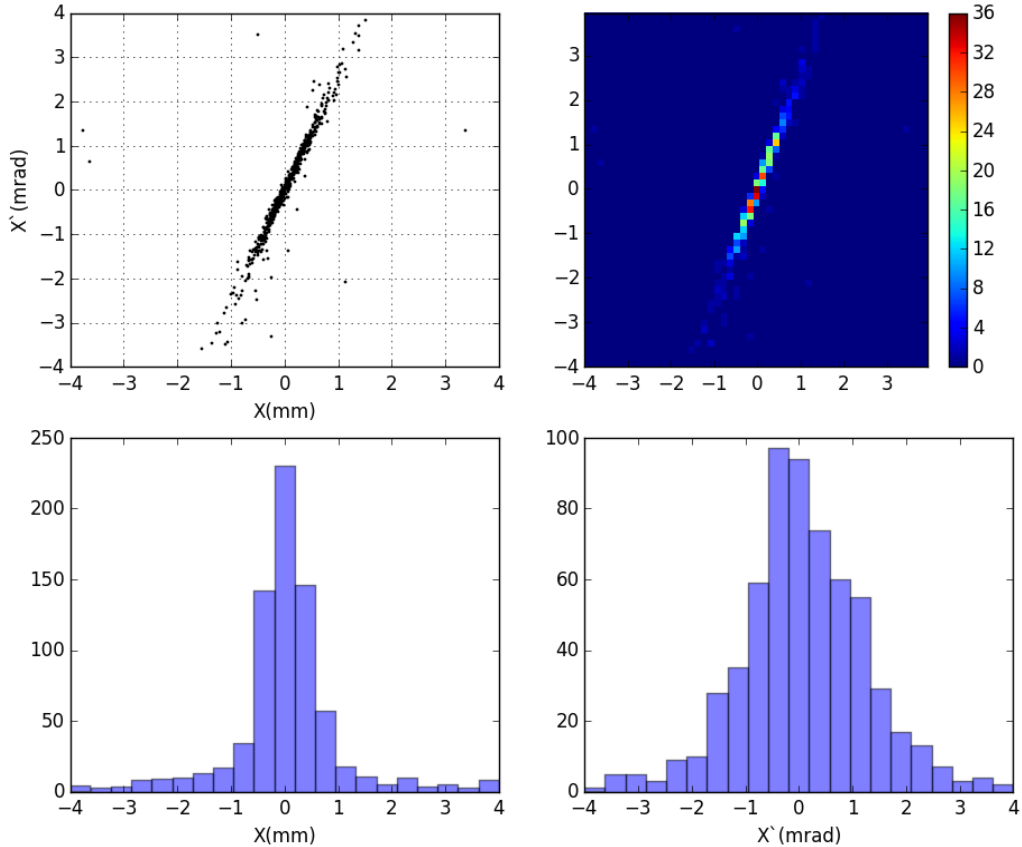

**Figure 8**: Diagrams of geometrical beam emittance for cooled beam with 1 µA beam current and 4 Pa $P_{RFQ}$: X-X' phase space distribution (top), transversal X-axis histogram plot (bottom-left), transversal X'-axis histogram plot (bottom-right).

## 4. Coupling RFQ–HRS: quadrupole triplet

The cooled beams outgoing from SHIRaC are accelerated toward an HRS entrance [24] where they should pass through a rectangular slit which has a dimension of 1×4 mm². Without any ion optical element, the ions losses are significant and less than 20 % of them can reach the HRS. In order to avoid these losses, the element should be installed between the extraction section exit and the HRS entrance and must allow focusing the beam at that slit. It also must be run with only a few kilovolts of voltage DC (less than 5 kV) for energy beams going up to 60 keV. The suitable solution is a multiplet of electrostatic quadrupole which can provide focus that is stronger than the Einzel lens [25].

Because the tuning of a quadrupole triplet is much easier than other quadrupole multiplet and because, at the same time, a triplet remains flexible enough to satisfy various requirements in the first order focusing properties, a quadrupole triplet have been adopted. The design of the triplet and the optimized dimensions of their quadrupoles are done by COSY INFINITY [26]. The optimum design is shown in figure 9 and their dimensions are illustrated in table 1.

Simulations have shown important aberrations of the beam through the triplet, due to the edge effects of their electrodes. In order to reduce them, collimators were installed on either side of each quadrupole.

Voltages are applied to the triplet in such a way that, in both x–z and y–z planes, converging and diverging lenses alternate. The y–z plane has been taken to be the DCD (diverging–converging–diverging) plane and x–z plane to be the CDC (converging–diverging–converging) plane. The optimum voltages for a beam of 5 keV are of 130 V, 130 V and 220 V respectively for the first, second and third quadrupoles. However, for a beam of 60 keV they are of 2.3 kV, 2.3 kV and 4 kV respectively.

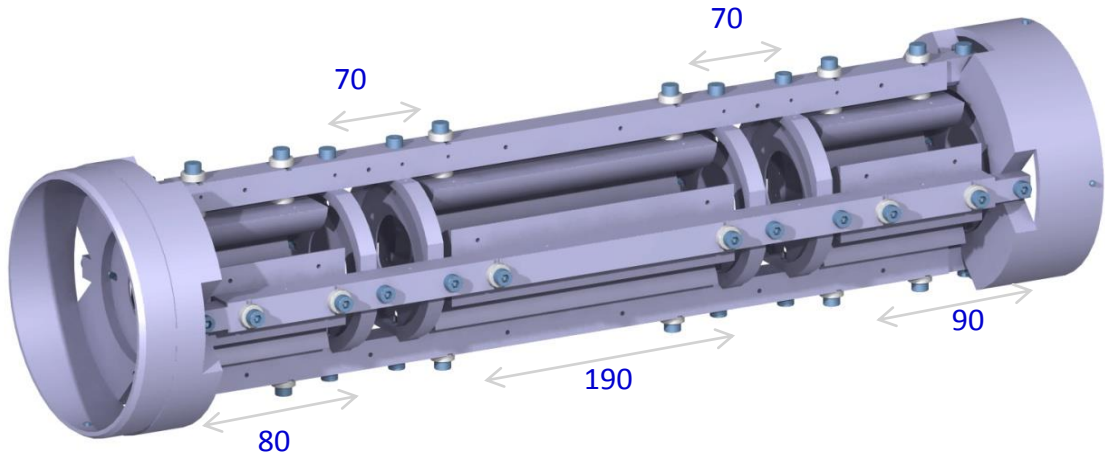

**Figure 9**: Schematic view of the electrostatic quadrupole triplet: three quadrupoles with collimator to each quadrupole's side.

| Compounds | Dimensions (mm) |
|---|---|
| 1st quadrupole (EQ 1) | 80 |
| Drift: EQ 1- EQ 2 | 60 |
| 2nd quadrupole (EQ 2) | 160 |
| Drift: EQ2-EQ 3 | 60 |
| 3rd quadrupole (EQ 3) | 80 |
| Drift: EQ 3 - HRS entrance | 230 |
| Total length | 440 |
| Inner radius | 15 |
| Length of the collimator | 2 |
| Inner radius of the collimator | 15 |

**Table 1**: Specification of the electrostatic quadrupole triplet.

In the below figure, we have presented the measurements of the transverse cooled beam profiles at the DCD and CDC planes for various beam currents. The space charge effect on the transverse size of the beams is clear while there are a widening of these profiles.

To quantitatively study this phenomenon, the FWHM ($\sigma_x$ or $\sigma_y$) of these profiles were determined, as presented in table 2. For beam currents going up to 1µA, the triple of these FWHM ($3.\sigma_x$ or $3.\sigma_y$) does not exceed the dimension of the slit ($3.\sigma_x < 1$ mm and $3.\sigma_y < 4$ mm). Therefore, more than 95% of cooled beams can pass through the slit toward the HRS.

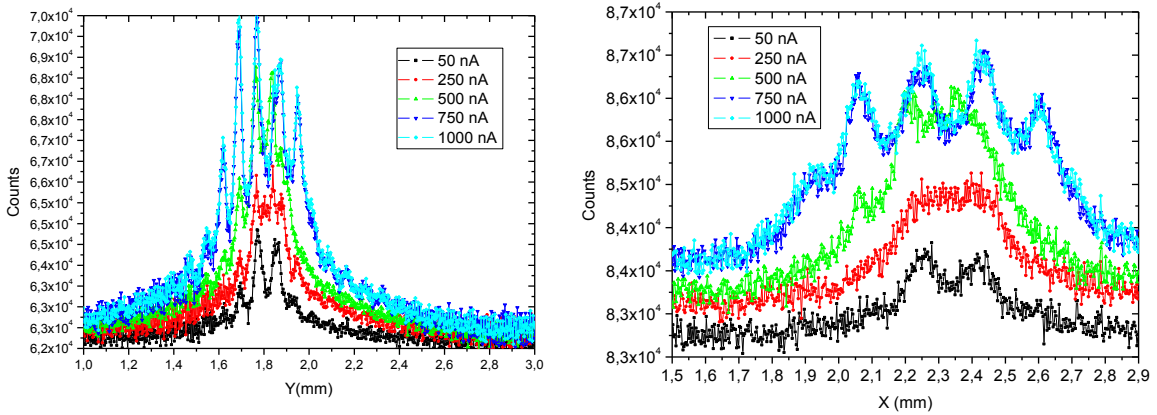

**Figure 10**: Transverse profiles of the cooled beams measured at the exit of the HRS slit for various beam currents and with optimum cooling condition showed above: profile at the CDC plan (left) and its counterpart at the DCD plan (right).

| Intensity (nA) | $\sigma_x$ (mm) | $\sigma_y$ (mm) |
|---|---|---|
| 50 | 0.231 | 0.347 |
| 250 | 0.246 | 0.350 |
| 500 | 0.242 | 0.362 |
| 750 | 0.308 | 0.579 |
| 1000 | 0.374 | 0.586 |

Table 2: The FWHM (Full half high Medium) of Gaussian fits of the cooled beam transverse profiles for various beam currents.

## 5. Nuclearization: confinement of nuclear matter

As mentioned above, SHIRaC will be implemented within the SPIRAL 2 line and will receive their beams. Because these beams will be of highest currents, going up to 1 µA, a high radioactivity environment into the experimental hall of SHIRaC would be very dangerous. To avoid this problem, specific considerations to confine the nuclear matter (radioactive chemical compounds) should be studied and developed. This requires, aside the buffer gas recycle, the usage of appropriate safety processes.

The confinement of the nuclear matter consists in avoiding the escape of any nuclear matter during the experimental test, mostly at the maintenance intervention. As the used vacuum system [16] ensures the statistic confinement of the nuclear matter, the dynamic confinement demands specific turbo molecular pumps with double valves and axial collar for the clamp (figure 11). The first valve is fixed to the module and the second one to the beam-line. The sweep of gussets between the double valves is performed before the disconnection of the module. The gas recovered following this sweep is analyzed and, if necessary, stored for radioactive decay.

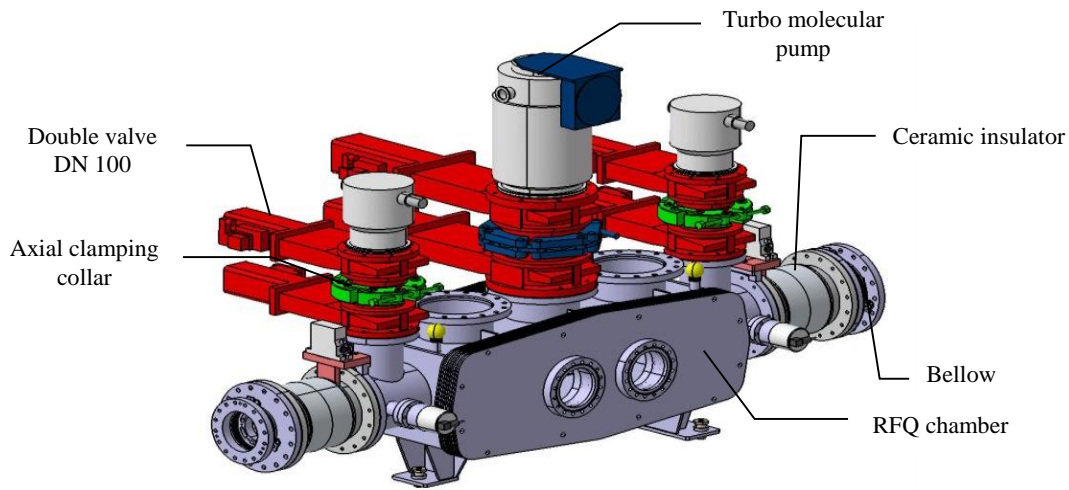

**Figure 11**: Technical design of the RFQ chamber and its interface: dynamic confinement method of the nuclear matter by turbo-molecular pumps with double valves and axial collar for clamp.

To maintain the RFQ module, withdraw or create elements under vinyl, the nuclear matter should be confined into a volume. This volume can occur via a confinement vinyl. Before opening the RFQ chamber, we should setup a waterproof sleeve which allows to avoid the escape of the nuclear matter. The steps of this process are done as follows (Figure 12): First, we set up a confinement vinyl. Then, in order to create two distinct volumes, we withdraw the RFQ support and, finally, we separate the two subsets. During the implementation phase of the RFQ, we should, foremost, setup the waterproof sleeve and then withdraw it before mounting the RFQ.

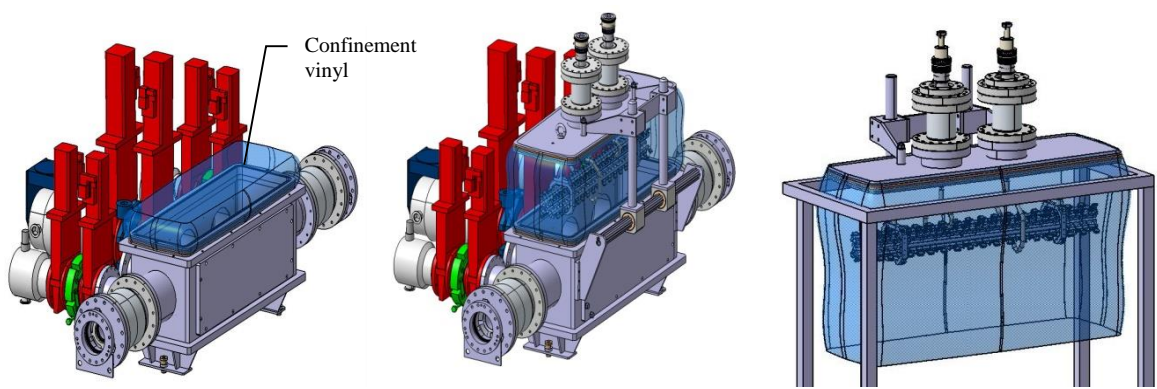

**Figure 12**: Steps of the RFQ withdraw: Establishment of confinement vinyl (left), withdraw and creation of two volumes (middle) and separation of two subsets (right).

# Conclusion and perspectives

The ingenious solution of the µRFQ has showed a high efficiency in enhancing the cooled beam quality and in reducing both the transverse geometric emittance and the longitudinal energy spread. The cooled beam quality reached the expected values of SHIRaC requirements and thereafter an isobaric purification with the HRS will be possible.

The proposed solutions for the nuclearization protection and the µRFQ presented in this paper are under realization.

Owing these studies, SHIRaC prototype will provide cooled beam quality never reached so far, and beam currents never handled before. However, the cooling of beam currents exceeding 1 µA will need a new design of RFQ Cooler and even the µRFQ will not be able to overcome the degrading effects. Special considerations of the extraction section should be studied for the next generation of RFQ Cooler for the EURISOL project [27, 28] where the beam current will reach 10 µA.

# Acknowledgment

We would like to thank Professor G.Ban for continuous encouragement and support. The vacuum system and mechanical design at L.P.C Caen (France) are gratefully acknowledged for their kind assistance in the development of the project. We also thank Mouna yahyaoui for her English corrections.